\documentclass[11pt,twoside]{article}
\usepackage{./asp2014}
\usepackage{comment}

\resetcounters

\begin{document}

\title{Radio Continuum Emission from Galaxies: An Accounting of Energetic Processes}
\author{Eric J. Murphy,$^1$ 
James J. Condon,$^1$
Antxon Alberdi,$^{2}$
Loreto Barcos-Mu{\~n}ozarcos,$^{3,1}$
Robert J. Beswick,$^{4}$
Elias Brinks,$^{5}$
Dillon Dong,$^{6}$
Aaron S. Evans,$^{7,1}$
Kelsey E. Johnson,$^{7,1}$
Rober C. Kennicutt Jr.,$^{8}$
Sean T. Linden,$^{7,1}$ 
Tom W. B. Muxlow,$^{4}$
Miguel P{\'e}rez-Torres,$^{2}$
Eva Schinnerer,$^{9}$
Mark T. Sargent,$^{10}$
Fatemeh S. Tabatabaei,$^{2,11}$ and 
Jean L. Turner$^{12}$
}
\affil{$^1$National Radio Astronomy Observatory, Charlottesville, VA 22903; \email{emurphy@nrao.edu}}
\affil{$^2$Instituto de Astrofisica de Andaluc\'{i}a (IAA, CSIC); Glorieta de la Astronom\'{i}a s/n, 18008-Granada, Spain}
\affil{$^3$Joint ALMA Observatory, Alonso de C\'{o}rdova 3107, Vitacura, Santiago, Chile}
\affil{$^4$Jodrell Bank Centre for Astrophysics/e-MERLIN, The University of Manchester, Manchester, M13 9PL, UK}
\affil{$^5$School of Physics, Astronomy and Mathematics, University of Hertfordshire, Hatfield AL10 9AB, UK}
\affil{$^6$California Institute of Technology, MC 100-22, Pasadena, CA 91125, USA}
\affil{$^7$Department of Astronomy, University of Virginia, 3530 McCormick Road,Charlottesville, VA 22904, USA}
\affil{$^8$Institute of Astronomy, University of Cambridge, Madingley Road, Cambridge CB3 0HA, UK}
\affil{$^{9}$Max Planck Institut f\"{u}r Astronomie, K\"{o}nigstuhl 17, Heidelberg D-69117, Germany}
\affil{$^{10}$Astronomy Centre, Department of Physics and Astronomy, University of Sussex, Brighton BN1 9QH, UK}
\affil{$^{11}$Departamento de Astrof\'{i}sica, Universidad de La Laguna, E-38206 La Laguna, Spain}
\affil{$^{12}$Department of Physics and Astronomy, UCLA, Los Angeles, CA 90095, USA}

\paperauthor{Eric J. Murphy}{emurphy@nrao.edu}{0000-0001-7089-7325}{National Radio Astronomy Observatory}{}{Charlottesville}{VA}{22903}{USA}
\paperauthor{James J. Condon}{jcondon@nrao.edu}{0000-0003-4724-1939}{National Radio Astronomy Observatory}{}{Charlottesville}{VA}{22903}{USA}

\begin{abstract}
Radio continuum observations have proven to be a workhorse in our understanding of the star formation process (i.e., stellar birth and death) from galaxies both in the nearby universe and out to the highest redshifts.  
In this article we focus on how the ngVLA will transform our understanding of star formation by enabling one to map and decompose the radio continuum emission from large, heterogeneous samples of nearby galaxies on $\gtrsim 10$\,pc scales to conduct a proper accounting of the energetic processes powering it.  
At the discussed sensitivity and angular resolution, the ngVLA will simultaneously be able to create maps of current star formation activity at $\sim$100\,pc scales, as well as detect and characterize (e.g., size, spectral shape, density, etc.) discrete H{\sc ii} regions and supernova remnants on 10\,pc scales in galaxies out to the distance of the Virgo cluster.  
Their properties can then be used to see how they relate to the local and global ISM and star formation conditions.  
Such investigations are essential for understanding the astrophysics of high-$z$ measurements of galaxies, allowing for proper modeling of galaxy formation and evolution.  

\end{abstract}

\section{Introduction}

Radio continuum observations have proven to be a workhorse in our understanding of the star formation process (i.e., stellar birth and death) from galaxies both in the nearby universe and out to the highest redshifts.  
A next-generation Very Large Array (ngVLA), as currently proposed, would revolutionize our understanding of the emission mechanisms that power the radio continuum emission in and around galaxies by enabling the routine construction of $\sim 1.2 - 116$\,GHz radio spectral maps. 
Coupled with its nearly order of magnitude increased sensitivity compared to the current Jansky VLA, the ngVLA makes it possible to use this frequency window to investigate the distinct physical processes linked to stellar birth and death for large, heterogeneous samples of galaxies for the first time.  
Furthermore, by delivering such a finely sampled spectrum over this entire frequency range with a single instrument will allow robust separation of the various emission components, which is currently the main challenge for multi-frequency radio studies.  
Each observation will provide enough sensitivity and spectral coverage to robustly decompose and accurately quantify the individual energetic components powering the radio continuum, thus providing unique information on the non-thermal plasma, ionized gas, and cold dust content in the disks and halos of galaxies.  


In this chapter we focus on how the ngVLA can be used to map and decompose the radio continuum emission from large heterogeneous samples of nearby galaxies on $\gtrsim 10$\,pc scales to conduct a proper accounting of their energetics.  
In doing so, we will be able to determine how star formation, AGN, and physical conditions in the ISM give rise to varying contributions of: 
\begin{itemize}
\item non-thermal synchrotron emission powered by accelerated CR electrons/positrons.
\item free-free emission from young massive star-forming (H{\sc ii}) regions.
\item anomalous microwave emission (AME), which is a dominant, but completely unconstrained, foreground in CMB experiments.  
\item cold, thermal dust emission that accounts for most of the dust and total mass content of the ISM in galaxies. 
\end{itemize}

\articlefigure{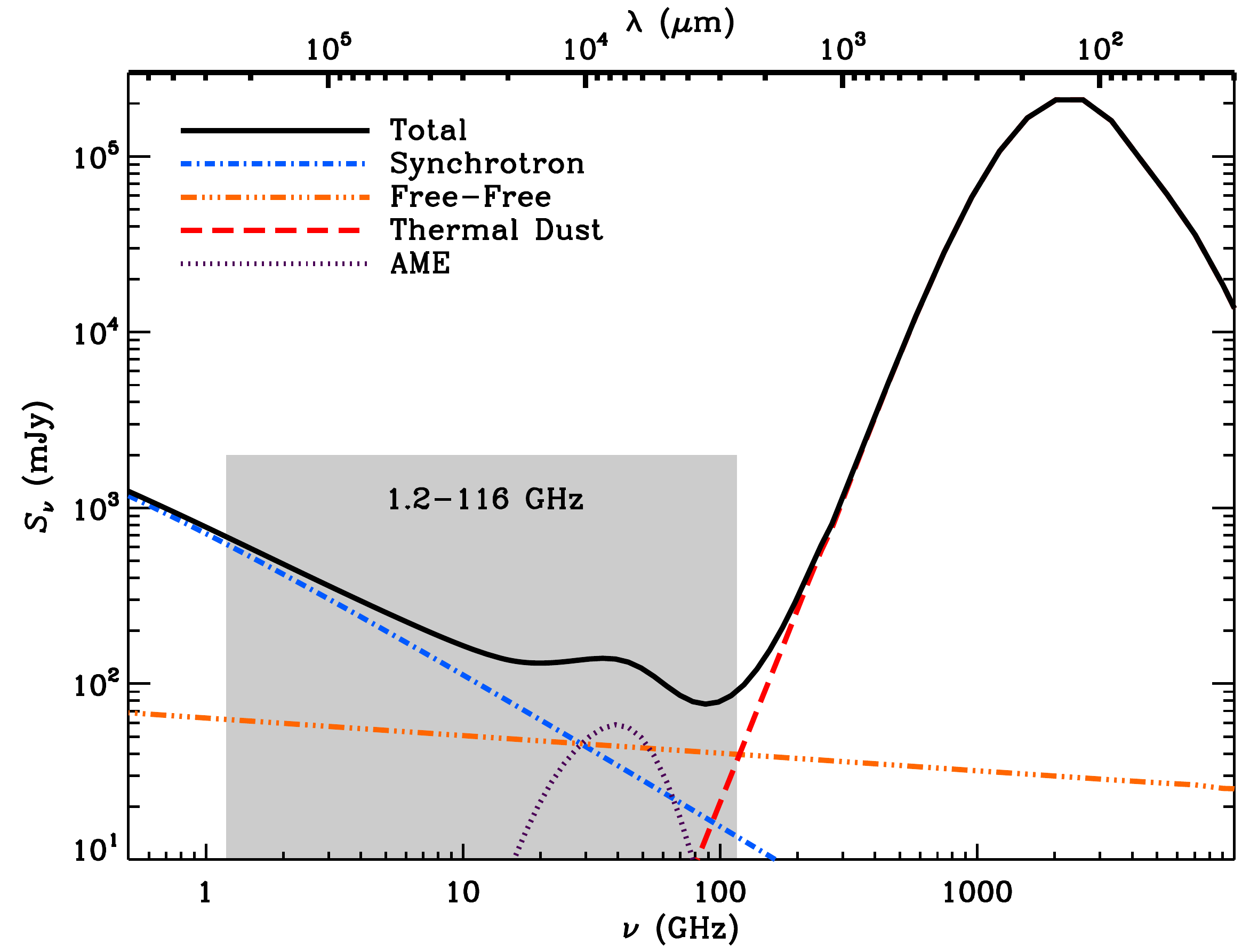}{fig:spec}{A model spectrum illustrating the various emission processes (non-thermal synchrotron, free-free, AME, thermal dust) that contribute to the observed radio frequency range to be covered by the ngVLA. Only in the proposed ngVLA frequency range ($1.2-116$\,GHz, highlighted) do all major continuum emission mechanisms contribute at similar levels, making this range uniquely well-suited to next-generation continuum studies.}

\section{The Energetic Processes Powering Radio Continuum Emission}
Radio continuum emission from galaxies covering $\sim1.2 - 116$\,GHz is powered by an eclectic mix of physical emission processes, each providing completely independent information on the star formation and ISM properties of galaxies (see Figure~\ref{fig:spec}).  
As such, it provides a unique window into the ISM and the process of star formation in galaxies.  
These processes include non-thermal synchrotron, free-free (thermal bremsstrahlung), anomalous microwave, and thermal dust emission that are directly related to the various phases of the ISM and provide a comprehensive picture of how galaxies convert their gas into stars.  
Each of these emission components, described in detail below, are of low surface brightness in the $\sim 30 -116$\,GHz frequency range, and therefore difficult to map in a spatially resolved manner at $\sim 10 - 100$\,pc scales in the general ISM of nearby galaxies using existing facilities.  
Consequently, our current knowledge about the emissions processes over this frequency range is limited to the brightest star-forming regions/nuclei in the most nearby sources \citep[e.g.,][]{mc08,mc10,akl11,ejm13,ejm15,lb15}, providing no information on how the situation may change in drastically different ISM conditions \citep[e.g.,][]{fat18} that may be more representative of those in high-redshift galaxies, where we have to rely on globally integrated measurements.   

\begin{itemize}
\item {\bf Non-Thermal Synchrotron Emission:} 
At $\sim$GHz frequencies, radio emission from galaxies is dominated by non-thermal synchrotron emission resulting, indirectly, from star formation. Stars more massive than $\sim$8\,$M_{\odot}$ end their lives as core-collapse supernovae, whose remnants are thought to be the primary accelerators of cosmic-ray (CR) electrons, giving rise to the diffuse synchrotron emission observed from star-forming galaxies. 
Thus, the synchrotron emission observed from galaxies provides a direct probe of the still barely understood relativistic (magnetic field + CRs) component of the ISM. 
As illustrated in Figure~\ref{fig:spec}, the synchrotron component has a steep spectral index, typically scaling as $S_{\nu} \propto \nu^{-0.83}$ with a measured rms scatter of 0.13 \citep{sn97}. 
By covering a frequency range spanning $1.2 - 116$\,GHz, the ngVLA will be sensitive to CR electrons spanning an order of magnitude in energy (i.e., $\sim1-30$\,GeV), including the population that may drive a dynamically-important CR-pressure term in galaxies \citep[e.g.,][]{as08}.
Further, the ngVLA will allow for investigations of the radio spectra of galaxies that are not self-similar across a range of physical scales as the injection of fresh CRs are able to modify the spectra locally.  

\item {\bf Free-Free Emission:}  
The same massive stars whose supernovae are directly tied to the production of synchrotron emission in star-forming galaxy disks are also responsible during their lifetime for the creation of H{\sc ii} regions. 
The ionized gas produces free-free emission, which is directly proportional to the production rate of ionizing (Lyman continuum) photons and optically-thin at radio frequencies. 
In contrast to optical recombination line emission, no hard-to-estimate attenuation term is required to link the free-free emission to ionizing photon rates, making it an ideal, and perhaps the most reliable, measure of the current star formation in galaxies.  
Unlike non-thermal synchrotron emission, free-free emission has a relatively flat spectral index, scaling as $S_{\nu} \propto \nu^{-0.1}$. 
Globally, free-free emission begins to dominate the total radio emission in normal star-forming galaxies at $\gtrsim$30\,GHz \citep[e.g.,][]{jc92, ejm12}, exactly the frequency range for which the ngVLA will be delivering an order of magnitude improvement compared to any current or planned facility.  

\item {\bf Thermal Dust Emission:}  
At frequencies $\gtrsim$100\,GHz, (cold) thermal dust emission on the Rayleigh-Jeans portion of the galaxy far-infrared/sub-millimeter spectral energy distribution can begin to take over as the dominant emission component for regions within normal star-forming galaxies. 
This in turn provides a secure handle on the cold dust content in galaxies, which dominates the total dust mass. 
For a fixed gas-to-dust ratio, the total dust mass can be used to infer a total ISM mass \citep[e.g.,][]{ch88,td01,nzs16}. 
Given the large instantaneous bandwidth offered by the ngVLA, approximately an order of magnitude increase in mapping speed at 100\,GHz compared to ALMA \citep{memo12}, such observations will simultaneously provide access to the $J=1\rightarrow0$ line of CO revealing the molecular gas fraction for entire disks of nearby galaxies. 
Alternatively, combining H{\sc i} observations (also obtained with the ngVLA) with $J=1\rightarrow0$ CO maps, one can instead use the thermal dust emission to measure the spatially varying gas-to-dust ratio directly.  

\item {\bf Anomalous Microwave Emission (AME):}  
In addition to the standard Galactic foreground components described above, an unknown component has been found to dominate over these at microwave frequencies between $\sim 10-90$\,GHz, and is seemingly correlated with 100\,$\mu$m thermal dust emission. 
Cosmic microwave background (CMB) experiments were the first to discover the presence of AME \citep{ak96,el97}, whose origin still remains unknown \citep[see][for a review]{cd18}. 
Its presence as a foreground is problematic as the precise characterization and separation of foregrounds remains a major challenge for current and future CMB experiments \citep[e.g.,][]{b215,planck16}. 
At present, the most widely accepted explanation for AME is the spinning dust model \citep{we57,dl98,planck11,hd17} in which rapidly rotating very small grains, having a nonzero electric dipole moment, produce the observed microwave emission. 
The increased sensitivity and mapping speed of the ngVLA will allow for an unprecedented investigation into the origin and prominence of this emission component both within our own galaxy and others, ultimately helping to improve upon the precision of future CMB experiments.    

\end{itemize}

\articlefiguretwo{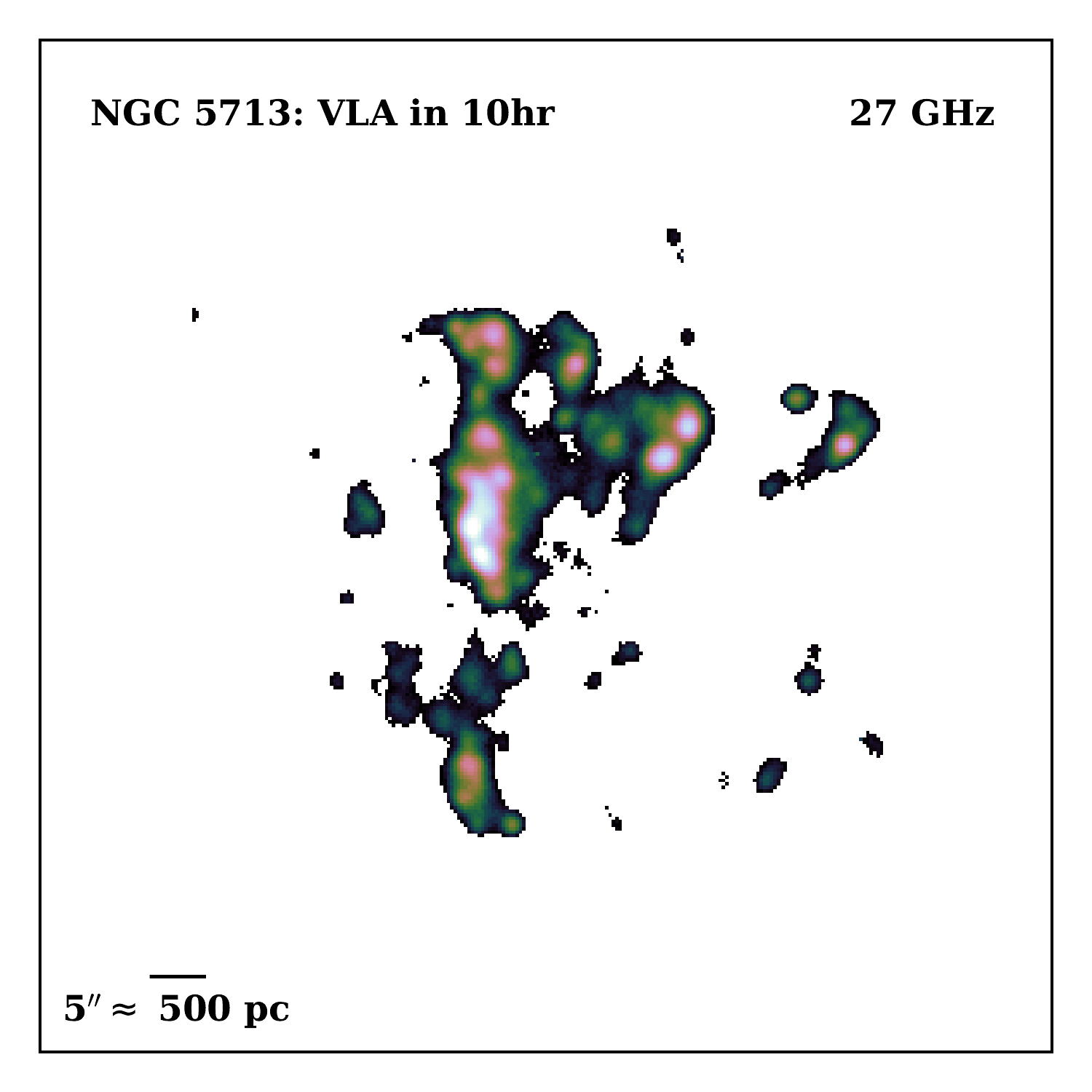}{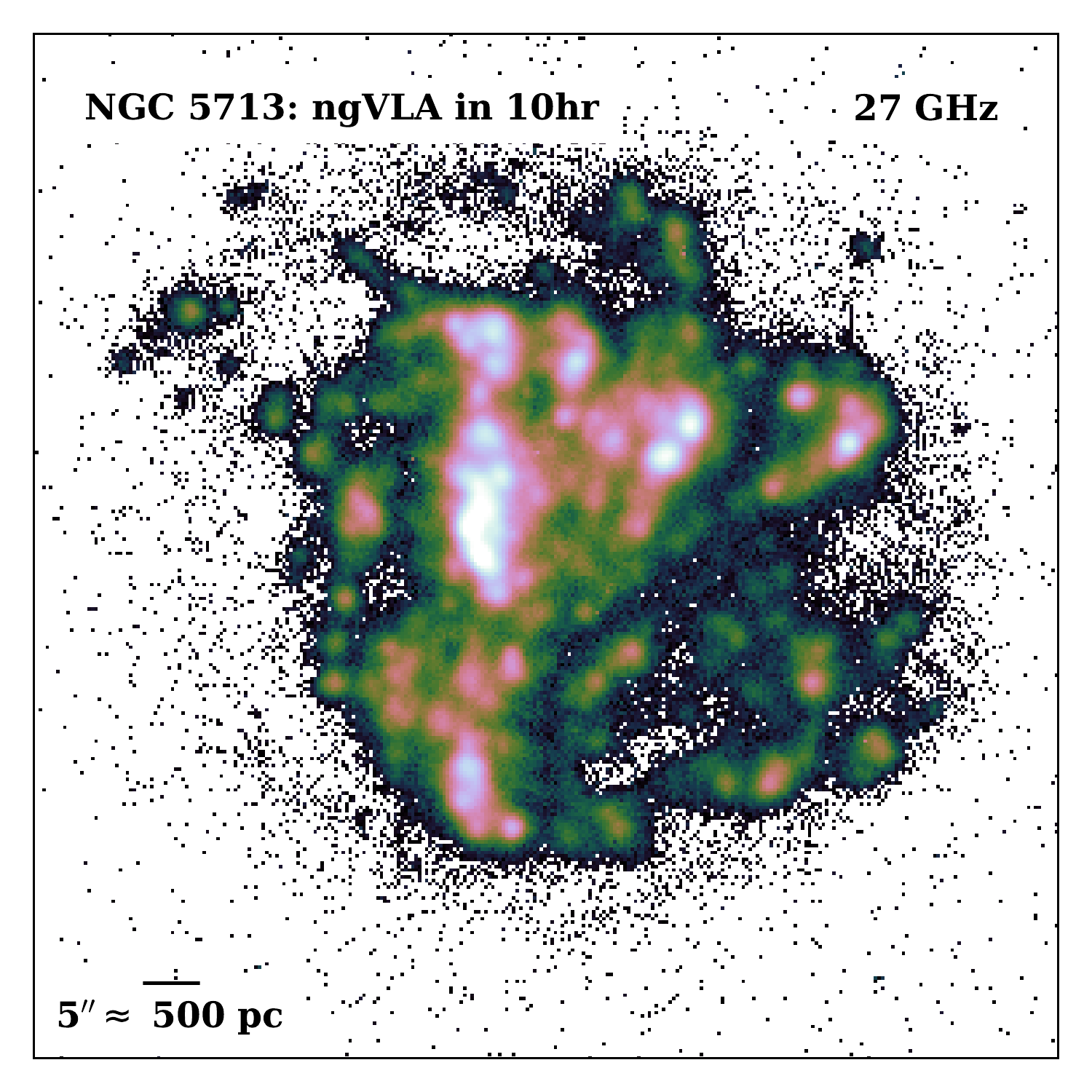}{fig:maps}{
Both panels show a model 27\,GHz free-free emission image of NGC\,5713  ($d _{L}\approx 21.4\,\mathrm{Mpc}, \mathrm{SFR} \approx 4\,M_{\odot}\,{\rm yr}^{-1}$) based on an existing H$\alpha$ narrow band image at its native ($\approx 1\arcsec$) angular resolution taken from SINGS \citep{rck03}.  
The left and right panels indicate the emission that would be detected at the 3$\sigma$ level after a 10\,hr on-source integration time using the current VLA  in C-configuration ($1\sigma \approx 1.5\, \mu{\rm Jy}\,{\rm bm}^{-1}$) and the ngVLA with a 1\arcsec~sculpted beam ($1\sigma \approx 0.17\, \mu{\rm Jy}\,{\rm bm}^{-1}$).  In a relatively modest integration, the ngVLA is able to recover a significant fraction of star formation activity that is completely missed by the VLA.  
}

\section{Robust Mapping of Star Formation within Nearby Galaxies on a Range of Physical Scales}
For a proper decomposition of the radio continuum emission into its component parts, one needs to have spectral coverage at frequencies low enough (i.e., $<10$\,GHz) to be dominated by the non-thermal, steep spectrum component, having a spectral index of $-0.83$ \citep{jc92,sn97,ejm11} and at frequencies high enough (i.e., $>50$\,GHz) where the emission becomes completely dominated by thermal emission, having a spectral index of $-0.1$.  
Given the potential for a significant contribution from AME \citep[e.g.,][]{ejm10,ejm18,as10,bh15}, peaking at frequencies $\sim20-40$\,GHz, coarse coverage spanning that spectral region is critical to account for such a feature.  
To date, the shape of the AME feature is largely unconstrained, and has not been carefully measured in the ISM of extragalactic sources.  
This is largely due to the insensitivity of current facilities to conduct a proper search for AME in nearby galaxies and map the feature with enough frequency resolution to provide useful constraints on its shape.  

Broadband imaging spanning the full ngVLA frequency range of $1.2 - 116$\,GHz will therefore be extremely powerful to properly decompose radio continuum emission from galaxies into its constituent parts.  
Additionally having frequency coverage below 8\,GHz provides sensitivity to free-free absorption, which is common in nearby luminous infrared galaxies \citep[e.g.,][]{jc91,mc10,lb15}.  
For individual (ultra compact) H{\sc ii} regions, the turnover frequency can be as high as $\approx$20\,GHz \citep{tm10}.  
A fundamental goal of the ngVLA will be to produce star formation maps for a large, heterogenous samples of nearby galaxies at $\approx$1\arcsec~resolution.  
This will deliver H$\alpha$-like images that optical astronomers have relied on heavily for decades without having the additional complications of extinction and contamination by nearby [N{\sc ii}] emission, which make such images challenging to interpret.       

By achieving arcsecond-like resolution that is commensurate with ground-based optical facilities, the ngVLA will be able to probe $\approx$100\,pc scales out to the distance of Virgo (the nearest massive cluster at $d \approx 16.6$\,Mpc), which are the typical sizes of giant molecular clouds (GMCs) and giant H{\sc ii} regions.  
At an rms sensitivity of 0.15\,$\mu$Jy\,bm$^{-1}$ at 27\,GHz, such radio maps will reach a sensitivity espressed in terms of star formation rate (SFR) density of $\approx 0.005\,M_{\odot}\,\mathrm{yr}^{-1}\mathrm{kpc}^{-2}$, matching the sensitivity of extremely deep H$\alpha$ images such as those included in the Local Volume Legacy survey \citep{rck08}.  
An example of this is illustrated in Figure~\ref{fig:maps} where an existing H$\alpha$ narrow band image taken from SINGS \citep{rck03} was used to create a model 27\,GHz free-free emission map at 1\arcsec~resolution for the nearby star-forming galaxy NGC\,5713.  
With a 10\,hr integration the ngVLA will be able to map the entire disk of NGC\,5713 down to an rms of $\approx$0.15\,$\mu$Jy\,bm$^{-1}$ ($\approx$35\,mK).  
A comparison of what can currently be delivered with the VLA for the same integration time is also shown, indicating that only the brightest star-forming peaks are able to be detected.     
To make a map to the same depth using the current VLA would take $\gtrsim$800\,hr!  
This is the same amount of time it would take to roughly survey $\gtrsim$80 galaxies with the ngVLA.

Using the same data, but applying a different imaging weighing scheme to create finer resolution maps (i.e., 0\farcs1, or even higher for brighter systems), similar multi-frequency radio continuum analyses can be performed for discrete H{\sc ii} regions and supernova remnants (SNRs) to complement high-resolution, space-based optical/NIR observations (e.g., {\it HST}, {\it JWST}, WFIRST, etc.).  
At an angular resolution of 0\farcs1, the data would sample $\approx$10\,pc scales in galaxies out to the distance of Virgo to resolve and characterize (e.g., size, spectral shape, density, etc.) discrete H{\sc ii} regions and SNRs with a sensitivity to diffuse free-free emission corresponding to a SFR density of $\approx 0.5\,M_{\odot}\,\mathrm{yr}^{-1}\mathrm{kpc}^{-2}$.  

This is a transformational step for studies of star formation in the local universe covering a large, heterogeneous set of astrophysical conditions. 
This statement is independent of the fact that with such observations using the ngVLA, having its wide-bandwidth, a number of RRL's will come for free \citep[e.g., see][in this volume]{db18}.
The detection of such lines (individually or through stacking), coupled with the observed continuum emission, can be used to quantify physical conditions for the H{\sc ii} regions such as electron temperature. 
It is without question that the ngVLA will make radio observations a critical component for investigations carried out by the entire astronomical community studying star formation and the ISM of nearby galaxies.

\articlefigure{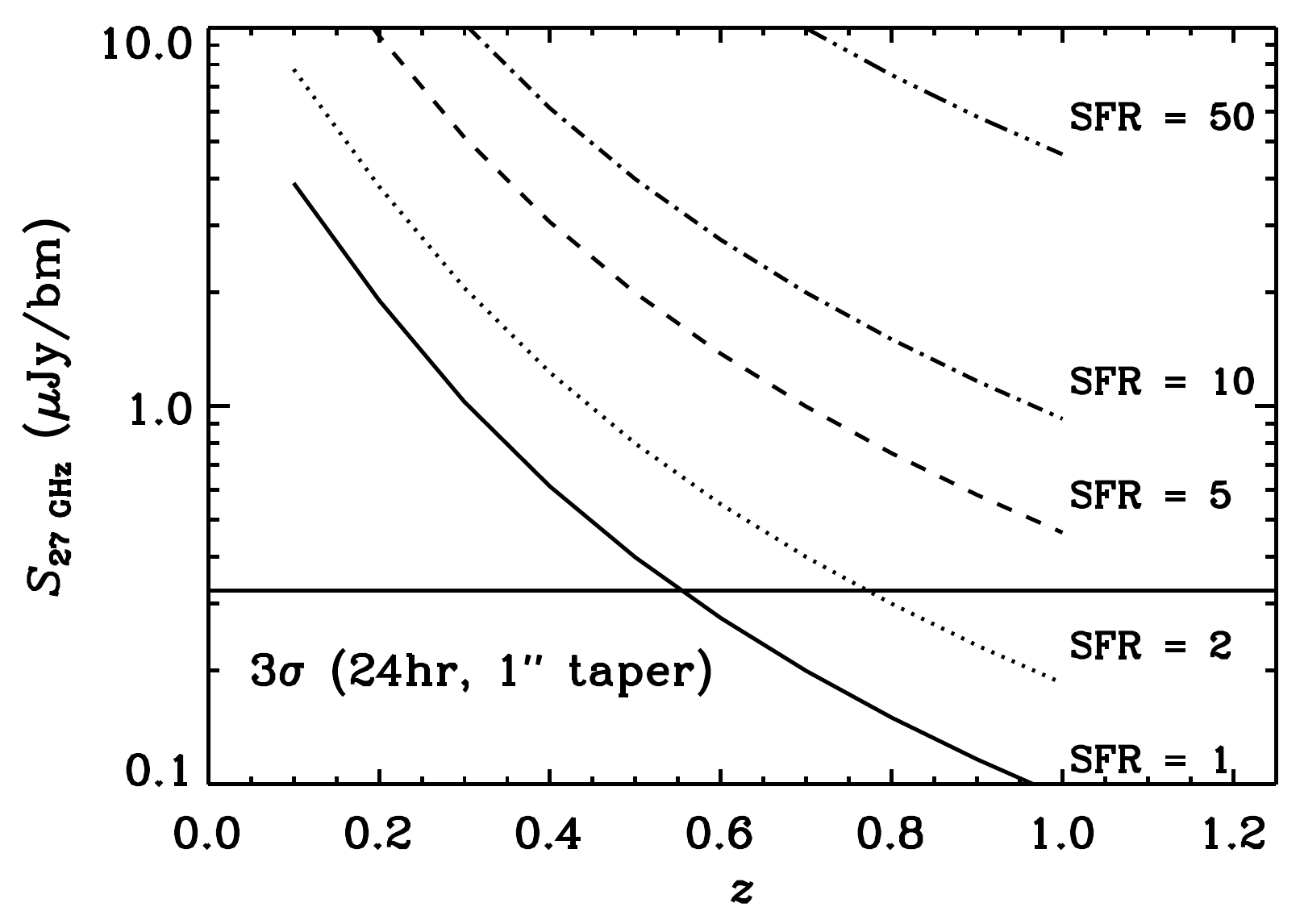}{fig:Tb_z}
{Telescope sensitivity in $\mu$Jy\,bm$^{-1}$ at 27\,GHz plotted against redshift indicating the expected brightness of a 4\,kpc disk galaxy forming stars at a rate of 1, 2, 5, 10, and 50\,$M_{\odot}\,\mathrm{yr}^{-1}$.   
Given the current sensitivity specifications, by tapering to a 1\arcsec~synthesized beam, the ngVLA will have enough brightness temperature sensitivity to resolve a Milky Way like galaxy forming stars at a rate of a $\sim 3\,M_{\odot}\,\mathrm{yr}^{-1}$ out to $z\sim1$ after a 24\,hr integration.   
Using the existing VLA, the detection of such a galaxy would take $\gtrsim$2000\,hr!}

\section{A Robust Star Formation Indicator at All Redshifts}
While deep field surveys aimed at studying the cosmic star formation activity at high redshifts is the focus of another chapter in this volume \citep[i.e., ][]{ab18}, we briefly describe what the ngVLA can do for studies of star formation out to moderate redshifts (i.e., $z \lesssim 1$).  
With the increased collecting area and bandwidth of the ngVLA, one would be able to measure and resolve rest-frame $60$\,GHz emission from a Milky Way like galaxy out to $z\sim1$ forming stars at rates of $\sim 3\,M_{\odot}\,\mathrm{yr}^{-1}$ after a 24\,hr integration. 
This is illustrated in Figure~\ref{fig:Tb_z} which shows the observed 27\,GHz brightness of 4\,kpc diameter disk galaxies with a range of SFRs. 
We note that a choice of 4\,kpc for a typical disk diameter may be slightly conservative, given that the 10\,GHz sizes found for $\mu$Jy sources at $z\sim1$ in deep VLA observations of GOODS-N are reported to be $\sim 1.2$\,kpc \citep{ejm17}.  
The 3\,$\sigma$ 27\,GHz rms of the ngVLA tapered to a 1\arcsec~synthesized beam is based on the updated sensitivities given in \citep{memo17}\footnote{\url{http://ngvla.nrao.edu/page/refdesign}}. 
Accordingly, such observations provide highly robust extinction-free measurements of SFRs for comparison with other optical/UV diagnostics to better understand how extinction on both galactic scales and within individual star-forming regions, evolves with redshift.  
By coupling these higher frequency observations with those at lower radio frequencies one can accurately measure radio spectral indices as a function of redshift to better characterize thermal and non-thermal energetics independently.  
This therefore enables one to determine if these physically distinct components remain in rough equilibrium with one another or if there are changes which in turn affects the global star formation activity as a function of lookback time.

\section{Uniqueness of ngVLA Capabilities}
While the VLA has frequency coverage up to 50\,GHz, it lacks the higher frequencies ($\gtrsim50$\,GHz) necessary to measure the location where the radio spectrum is completely dominated by free-free emission and/or properly constrain the spectral shape of AME.  
This is required to robustly account for the energetic contributions of each emission component to the total radio continuum emission of star-forming galaxies.  
Furthermore, at the top end of the ngVLA frequency range, cold thermal dust emission associated with the molecular ISM may also be detectable at low surface brightness levels, which can be characterized by a rising spectrum.  
Perhaps most importantly, the lack of sensitivity by the VLA and other telescopes (e.g., ALMA, NOEMA, etc.) makes such continuum mapping for entire galaxies, rather than just the brightest H{\sc ii} regions within them, impossible, requiring the factor of $\gtrsim 10\times$ improvement in sensitivity afforded by the ngVLA.  
It is only with the ngVLA that the full potential of the radio continuum spectrum can be used as a tool to properly constrain the various energetic processes that power its emission for a broad range of heterogeneous conditions in galaxies.  

\section{Synergies at Other Wavelengths}
Nearby galaxies provide our only laboratory for understanding the detailed physics of star formation and AGN activity across much larger ranges of physical conditions than offered by our own Milky Way.  
They are the workhorses for testing and applying physical models of star formation and their associated feedback processes that are used in galaxy evolution models to explain statistical observations for large populations of galaxies at high redshifts, for which it is impossible to conduct the detailed astrophysical experiments described here.  
Consequently, the transformational set of observations enabled by the ngVLA, and discussed in this chapter, will illuminate the relation between thermal and non-thermal energetic processes in the ISM and be highly synergistic with observations from shorter wavelength ground- and space-based telescopes that have access to large numbers of other diagnostic (line and continuum imaging) that may be difficult to interpret due to extinction at both low and high redshift.  
The same can be said for synergy with longer wavelength, far-infrared telescopes that will provide access to dust continuum and fine structure line emission that can be used to characterize the cold/warm neutral phase of the ISM, but lack the angular resolution necessary to study discrete ($\gtrsim 10$\,pc) star-forming regions in large samples of galaxies.  





\end{document}